# Transformer for seismic image super-resolution

Shiqi Dong[1], Xintong Dong[2], Kaiyuan Zheng[1], Ming Cheng[2], Tie Zhong[1], Hongzhou Wang[2]

[1]Key Laboratory of Modern Power System Simulation and Control and Renewable Energy Technology (Ministry of Education), and Department of Communication Engineering, Northeast Electric Power University, Jilin, China.

[2]College of Instrumentation and Electrical Engineering, Jilin University, Changchun, China.

Shiqi Dong (e-mail: dsq1994@126.com); Xintong Dong (e-mail: dxt@jlu.edu.cn); Kaiyuan Zheng (e-mail: 1211756483@qq.com); Ming Cheng (e-mail: 2232058329@qq.com); Tie Zhong (e-mail: 519647817@qq.com); Hongzhou Wang (e-mail: hzwang21@mails.jlu.edu.cn)

ABSTRACT

Seismic images obtained by stacking or migration are usually characterized as low signal-to-noise ratio (SNR), low dominant frequency and sparse sampling both in depth (or time) and offset dimensions. For improving the resolution of seismic images, we proposed a deep learning-based method to achieve super-resolution (SR) in only one step, which means performing the denoising, interpolation and frequency extrapolation at the same time. We design a seismic image super-resolution Transformer (SIST) to extract and fuse local and global features, which focuses more on the energy and extension shapes of effective events (horizons, folds and faults, etc.) from noisy seismic images. We extract the edge images of input images by Canny algorithm as masks to generate the input data with double channels, which improves the amplitude preservation and reduces the interference of noises. The residual groups containing Swin-Transformer blocks and residual connections consist of the backbone of SIST, which extract the global features in a window with preset size and decrease computational cost meanwhile. The pixel shuffle layers are used to up-sample the output feature maps from the backbone to improve the edges, meanwhile up-sampling the input data through a skip connection to enhance the amplitude preservation of the final images especially for clarifying weak events. 3-dimensional synthetic seismic volumes with complex geological structures are created, and the amplitudes of half of the volumes are mixtures of strong and weak, then select 2-dimensional slices randomly to generate training datasets which fits field data well to perform supervised learning. Both numerical tests on synthetic and field data in different exploration regions demonstrate the feasibility of our method.

INTRODUCTION

The accuracy of seismic interpretation rely on seismic images with high resolutions, especially for some important geological targets like faults, folds, horizons, reservoirs and salt-dome (Miller et al., 1998; Tingdahl and Rooij, 2005). Due to the contamination of noises, attenuation of waves during propagation especially for high frequencies, physical limitation of sensors, and sparse acquisition geometry limited by environment or cost make the seismic images show characteristics of low signal-to-noise (SNR), low dominant frequency and sparse sampling both in depth (or time) and offset dimensions. The researches on improving the resolution of seismic images can be divided into three groups: data acquisition, prestack processing and processes on seismic images directly.

In terms of data acquisition, improving the acquisition density (Xiao et al., 2014), widening acquisition azimuth (Xiao et al., 2014; Li and Cao, 2017) and broadening frequency bandwidth of source signals excited by vibrators (Zeng et al. 2020) are efficient approaches to improve the resolution of seismic images. The dense acquisition requires more sources and receivers to make the intervals between both inlines and crosslines smaller for achieving better resolution in horizon. The wide acquisition azimuth requires larger coverage area of acquisition geometry to improve the vertical resolution of target formations. The sources with broadband signals improve the vertical resolution of seismic images as well. However, more sources and receivers, and wider acquisition azimuths increase the cost of acquisition dramatically, especially when using the expensive broadband sources and receivers. In addition, the huge amount of data obtained by dense acquisition greatly increases the cost of data processing.

A series of processes on prestack seismic data including denoising (Dong et al., 2019; Wang et al., 2022a; Zhong et al., 2022; Dong et al., 2023), interpolation (Niu et al., 2023; Dong et al.,

2024b), and attenuation compensation (Aharchaou and Neumann, 2019) may improve the quality of seismic images. Eliminating noises from field data improves SNR, but potentially eliminating part of the reflection events as well (Dong et al., 2022). Interpolation and attenuation compensation for prestack seismic data improve the lateral resolution and deep imaging, but inaccurate interpolation and attenuation compensation may introduce artificial noises (Chen and Yang, 2022; Cheng et al., 2023). Thus, improving the resolution of seismic images (including poststack profiles and migrated images) directly is a potential and important approach.

Processes on seismic images to improve resolution including denoising (Gorszczyk et al., 2015; Ketelhodt et al., 2019), interpolation (Chen et al., 2023; Li et al., 2024b), attenuation compensation (Zhou et al., 2022; Li et al., 2024a), deconvolution (Sui and Ma, 2020; Liu et al., 2022) and domain transformation (Alaei et al., 2018; Gremillion et al., 2018). Denoising the residual noises that are not suppressed on the pre-stack field data improves the SNR of seismic images. The attenuation compensation enhances the weak amplitudes generated from deep formations. The deconvolution and domain transformation methods extend the frequency bandwidth of traces in seismic images, and improve the peak frequency of traces to achieve enhancement of vertical resolution. Different methods play different roles in enhancing the resolution of seismic images. Thus, in order to obtain a seismic image with high-resolution, these methods or some of them need to be used in sequence. Multi-step processing increases the cost of computation, and the errors after each step are accumulated as the processing progresses (Zhong et al., 2023). Therefore, one-step and intelligent methods for super-resolution of seismic images are essential.

The development of GPU and deep learning algorithms provide chances to achieve complex nonlinear mapping, and has been widely used is many fields of geophysics, such as first break picking (Duan and Zhang, 2020), frequency extrapolation (Sun et al., 2023; Dong et al., 2024a), trace interpolation (Chai et al., 2020), denoising (Zhang et al., 2019), impedance inversion (Wang et al., 2022b), velocity inversion (Yang and Ma, 2023; Wang et al., 2023) and seismic interpretation (Wu et al., 2019). Dong et al. (2024c) used deep learning to compensate for the low imaging accuracy caused by sparse shots acquisition which improves the resolution of seismic images on the premise of saving acquisition costs, and provides a new perspective for improving the resolution of seismic images. In recent years, the research on super-resolution (SR) of seismic images based on deep learning has made some achievements as well. The networks used for seismic image SR including generative adversarial networks (GAN) (Oliveira et al., 2019; Sun et al., 2022; Liu et al., 2023; Lin et al., 2024), convolutional neural network (CNN) (Meng et al., 2021; Zeng et al., 2023; Zhou et al., 2023), U-Net (Li et al., 2021; Li et al., 2022; Hamida et al., 2023; Min et al., 2023; Gao et al., 2023) and diffusion model (Xiao et al., 2024). The prior information including faults (Zhou et al., 2023), facies (Hamida et al., 2023), binarized edge information (Min et al., 2023) and well-log data (Gao et al., 2023) can be used as the constraint to improve the results of SR. The training scheme used in most of the published works is supervised learning, but there is a weakly supervised method using unpaired low-resolution (LR) and high-resolution (HR) data trained by cycle GAN (Liu et al., 2023).

There are key points on deep-learning based SR of seismic images that need to be concerned. In general, supervised learning using synthetic datasets that are similar enough to the feature distributions of field data is better than unsupervised or semi-supervised learning, but the generalization of supervised ways usually not as good as that of unsupervised ways. In addition, the predicted HR images should be clearer and with higher SNR than LR images, whereas the amplitudes of the events in raw images should be preserved and fake geological structures should not be introduced. As for deep learning based methods, the task of SR for seismic images should not only need to pay attention to the resolution of local events, but also to the global information such as the shapes of horizons and the dip angle of faults whose features can not be well extracted only by using convolution-based networks.

Seismic images obtained after stacking or migration show obvious characteristics of texture which is similar to natural images, but the structure distributions of horizons and faults in the subsurface semi-space are special concerns of seismic images which determines the reliability of the SR results. Inspired by the powerful capability of global feature extraction within Transformer model (Vaswani et al., 2017), especially the Shifted-Window Transformer (Swin-Transformer) that is more suitable for vision and image tasks (Liu et al., 2021), which is widely used in SR for natural images (Liang et al., 2021), we propose a seismic image super-resolution Transformer (SIST) model and adopt supervised learning to achieve better SR for seismic images. For improving the anti-noise capability of the SR model and enriching the information of interfaces, we use Canny algorithm (Canny, 1986) to extract the binarized edge images of the input LR images as masks, which are used to be multiplied by the input images to generate the edge images with true amplitudes to compose the 2-channel input data. The Swin-Transformer layers with residual connections consist of the backbone of SIST, which extract and fuse the global and local features in shifted windows with a preset size, and save computational cost as well. The pixel shuffles are used to up-sample feature maps to obtain HR images with clearer edges, which act on both the output of backbone and original input data to improve the amplitude preservation of SIST especially for deep and weak events. We train SIST in the supervised approach, and we make 3-dimensional (3D) synthetic seismic volumes with complex geological structures and select 2-dimentsinal (2D) slices randomly to generate training datasets. In order to lead the training datasets fit field data better, we make the amplitudes of half of the slices are mixtures of strong and weak to improve the amplitude preserving further. We test the behaviors of well trained SIST on both synthetic (testing datasets and Marmousi model) and field data (Netherlands F3 block (dGB Earth Sciences, 1987), New Zealand Parihaka 3D survey (New Zealand Crown Minerals, 1996) and the North Sea Volve field (The Volve Data Village, 1993) to demonstrate the superior SR feasibility of the proposed method including amplitude preservation, strong generalization, better anti-noise capability and recovery of weak events.

## METHODS

**Problem definition**

Obtaining images with high SNR, denser sampling rate and clear (or thinner) events are the ultimate purpose of SR process for seismic images. Thus, traditional SR processes can be expresses as

$$\mathbf{I}_{HR} = Extp\left(Intp_{tod}\left(Intp_{dist}\left(D\left(\mathbf{I}_{LR}\right)\right)\right)\right) + R_{SR}\left(R_{D}, R_{Intp}, R_{Extp}; \Theta\right), \quad (1)$$

where $\mathbf{I}_{HR}$ and $\mathbf{I}_{LR}$ represent the seismic images with HR and LR, respectively. $D$ represents the denoising operator. $Intp_{dist}$ and $Intp_{tod}$ represent the interpolation operator in the dimension of distance and time (or depth), respectively. $Extp$ represents the operator of frequency extrapolation, and with regards to SR tasks, $Extp$ enhances the dominant frequency of LR images. $R_{SR}$ represents the total residuals of the final HR images after SR process, which is an accumulated residuals of denoising ($R_D$), interpolation ($R_{Intp}$) and extrapolation ($R_{Extp}$) due to the traditional SR perform each processing step in turn, and the $R_{SR}$ is affected by the parameters ($\Theta$) used in each traditional processing step. In addition, the multi-step processing incurs significant computational costs. Thus, reducing the processing steps and the manual selection of parameters will play an important role in improving the result of SR and decreasing residuals. Deep learning is an effective operator that achieves strong non-linear mapping between input data and target data in just one step

$$\mathbf{I}_{HR} = Net(\mathbf{I}_{LR};\theta) + R_{SR}, \tag{2}$$

where *Net* represents the artificial neural network, and $\theta$ denotes the trainable parameters of the network. Thus, deep learning methods based on reasonable and well-featured datasets, appropriate training approaches and effective architectures of networks are able to achieve efficient and accurate SR tasks.

**Architecture of SIST and its main blocks**

The architecture of SIST is tripartite as shown in Figure 1: input, backbone and reconstruction. For the input module, the binarized edge images of raw LR images are obtained by Canny algorithm. However, the binary images can only display the positions and shapes of the edges, but do not carry the amplitude information of the edges to reveal the contrasted relationship among different geological interfaces. Thus, we use the edge image as a mask to multiply with the raw image in elemental-wise (Eltwise) to generate the edge constraint with true amplitudes

$$\mathbf{I}_{edge} = Canny(\mathbf{I}_{LR}) \odot \mathbf{I}_{LR}, \tag{3}$$

where $\mathbf{I}_{edge}$ represents the edge image, and *Canny* represents the Canny operator. $\odot$ denotes the calculation of elemental-wised multiplication. $\mathbf{I}_{edge}$ is concatenated (Concat) with the raw image ($\mathbf{I}_{LR}$) to generate the input data with two channels. The input data enters the backbone through a convolutional layer with 3×3 filters.

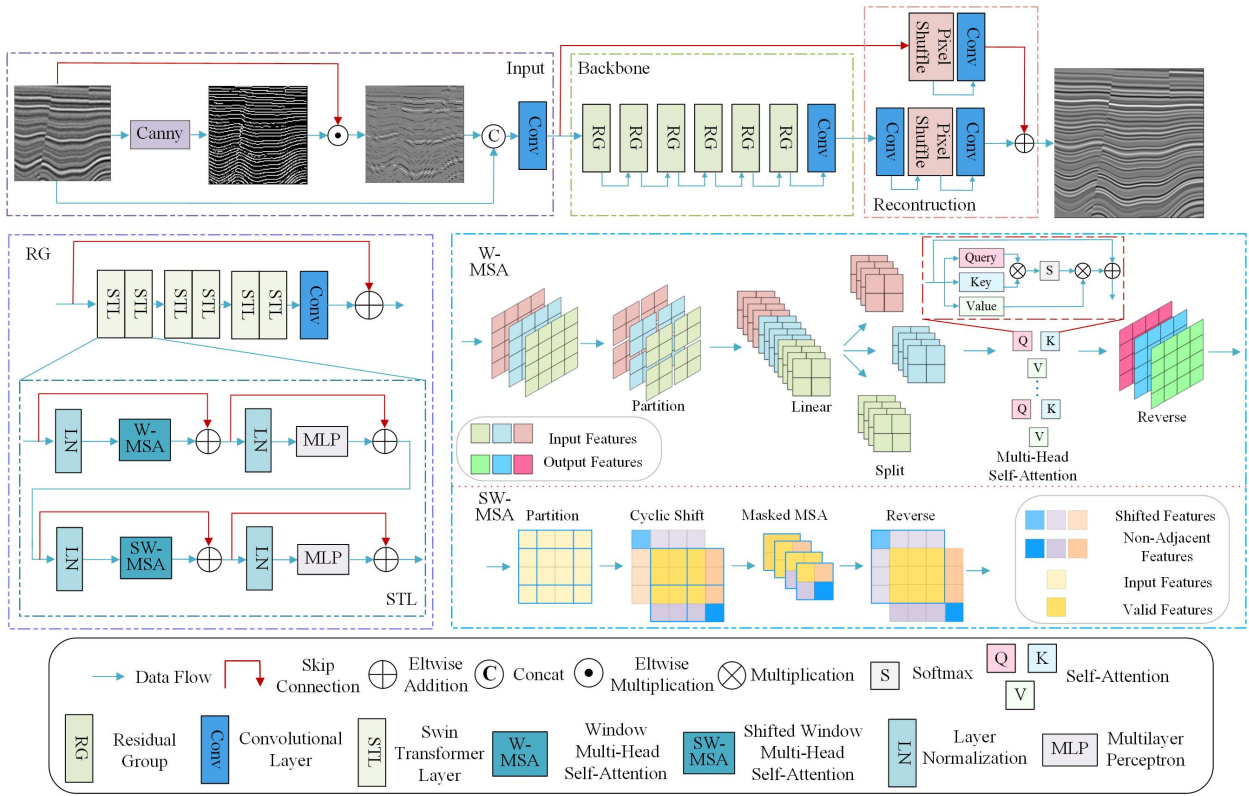

**Figure 1. The architecture of SIST and its main blocks.**

The backbone plays the role of Swin-Transformer and is a key step in enabling local and global feature extraction and fusion of input data, which employs 6 residual groups (RG) and a convolutional layer with 1×1 filters connected in sequence. In order to improve the convergence and stability of training for deep networks, and save computational cost at the same time, RG consists of three groups of Swin-Transformer layers (STL) and a convolution layer with 3×3 filters, and the input data and output feature maps of RG are Eltwise added by a skip connection to form the residual connection. Each group of STL consists of two connected STL. For the first STL in a group of STL, the input feature maps are layer normalized and calculated by a windows multi-head self-attention (W-MSA) block. In the W-MSA block, the input feature maps in each channel is partitioned by non-overlapped windows with a same preset size, and the feature maps inside a window are linearized and split in channel dimension then calculated by the multi-head

self-attention (MSA), finally the features are reverse to the same dimension of the input data

$$\mathbf{F}_{out} = Rev\left(MSA\left(Spl\left(Lin\left(Part\left(\mathbf{F}_{in};W\right);C\right)\right);H\right)\right), \quad (4)$$

where $\mathbf{F}_{in}$ and $\mathbf{F}_{out}$ represent the input and output feature maps. *Part*, *Lin*, *Spl*, *MSA* and *Rev* represent the operator of partition, linearize, split, MSA and reverse, respectively. *W*, *C* and *H* represent the preset windows, channels and the number of heads used for calculating self-attention (SA). In the SA block, the features in each channel after split can be divided to three feature maps: Query (**Q**), Key (**K**) and Value (**V**) by linear transformation to calculate the self-attention

$$\mathbf{F}_{out} = \left[soft\max\left(\frac{\mathbf{Q}\times\mathbf{K}^T}{\sqrt{d_k}}\right)\times\mathbf{V}\right] + \mathbf{F}_{in}, \quad (5)$$

where $d_k$ represents the dimension of **Q** and **K** to avoid vanishing gradient and exploding gradient problems. *T* denotes the transposition. The concatenation of SA for each head in channel dimension composes the output of MSA. SA is the method for obtaining global features generally, but in W-MSA the SA is calculated in a window which extracts the local features. Thus, W-MSA enables SIST to pay attention to the local features of seismic images, such as local relationships between signals and noises as well as the amplitudes of weak events. For extracting the features in cross-windows to obtain global features, the second STL in a group of STL employs shifted window based MSA (SW-MSA). As shown in Figure 1, for the input features of SW-MSA block, the window shifted one grid to the lower right aims to calculate W-MSA in a cross-window which makes the partitioned input data uneven. Thus, shifted the features from their original positions to the opposite end of the input feature maps to make the input features be partitioned evenly by non-overlapped windows. However, there is no spatial adjacency between the shifted and intact features, which makes no sense to calculate MSA in a windows contains mixture of shifted and intact features. Therefore, non-adjacent features are regarded as invalid data and intact features are valid data to generate masks, so that only the MSA among pixels of intact features in a window can be calculated. Finally shifting these shifted features reversely to their original positions which prepares for the next shift to achieve cyclic shift. The data transformation in SW-MSA block can be expresses as

$$\mathbf{F}_{out} = Rev\left(MMSA\left(Shift\left(Part\left(\mathbf{F}_{in};W,S,D\right)\right)\right)\right), \tag{6}$$

where *Shift* and *MMSA* represent the operator of cyclic shift and masked MS, respectively. *S* and *D* represent the shifted step size and direction, respectively. Under the process of SW-MSA, the preset windows extract the features in different positions of the input feature maps to implement the fusion of local and global features, which enables SIST pay attention to the shapes of geological structures in larger space as well. The output feature maps of 6 RG are through a convolutional layer with 1×1 filters to adjust the number of channels.

The image has basically obtained essentially effective features through the backbone, and then the feature maps need to be reconstructed to obtain the final HR images. We use the pixel shuffle layer (Shi et al., 2016) to implement up-sampling to avoid losing any useful information and improve the sharpness of edges in output images. The input data of the pixel shuffle layer can be arranged along the channel dimension

$$\mathbf{F}_{in}(h,w,C) = \left\{\mathbf{F}_1^1,\mathbf{F}_1^2,\cdots,\mathbf{F}_1^i,\cdots,\mathbf{F}_1^{h\times w};\mathbf{F}_2^1,\cdots;\mathbf{F}_j^1,\cdots;\mathbf{F}_C^1,\cdots,\mathbf{F}_C^{h\times w}\right\}_C, i \in [1,h\times w], j \in [1,C] \tag{7}$$

where *h* and *w* represents the height and width of the input feature maps in each channel. The output data of the pixel shuffle layer can be expressed as

$$\mathbf{F}_{out}\left(\sqrt{C}\times h,\sqrt{C}\times w\right)=PS\left(\mathbf{F}_{in};C\right)=\begin{bmatrix} \mathbf{F}_1^1 & \mathbf{F}_2^1 & \cdots & \mathbf{F}_{\sqrt{C}}^1 & \mathbf{F}_1^2 & \cdots \\ \vdots & \ddots & & \vdots & & \\ \mathbf{F}_{C-\sqrt{C}+1}^1 & \cdots & \ddots & \mathbf{F}_C^1 & \cdots & \\ & & & & \vdots & \\ \vdots & & & & \ddots & \\ & & \cdots & & & \mathbf{F}_C^{h\times w} \end{bmatrix} \quad (8)$$

where *PS* represents the operator of pixel shuffle layer. *PS* rearranges the elements in the same position in each channel of the input feature maps into a square in the output data. Therefore, the feature maps of different channels which carry different features of the input images compose the pixels in the same position in the output image to improve the resolution. In addition, the amplitudes of the raw images may be destroyed after complex feature extraction by backbone. Thus, we up-sample the raw two-channeled input data by *PS* through a skip connection and then add the two output images of *PS* to improve the amplitude preservation of the final HR images.

**Generation of datasets**

We train SIST in the supervised way which requires the training datasets that are similar to the field data in characteristic distribution. We follow the method (Li et al., 2022; Min et al., 2023) for generating 3D post-stack volumes which consists of three steps:

(1) Generating two groups of 3D reflectivity volumes involve horizons, folds, faults with the number of samples in each dimension are 128 and 256, respectively.

(2) Convolving the reflectivity volumes and Ricker wavelet with different peak frequencies to obtain the synthetic 3D post-stack data. We use the Ricker wavelet with a random peak frequency selected between 10~25 Hz to convolve the volumes with 128 side length to generate LR volumes, and the Ricker wavelet with a random peak frequency selected between 35~55 Hz to convolve the volumes with 256 side length to generate HR volumes. In addition, the LR volumes are added colored noises which are similar to the post-stack to decrease its SNR.

(3) Randomly pick two slices from a volume on the surface with two orthogonal directions. There are 3000 pairs of LR and HR 2D synthetic seismic images in total to compose the datasets. 80%, 10% and 10% of the datasets are used for training, validation and testing, respectively, and there is no over-lapping among these sub-datasets.

Due to the distribution of the subsurface reflectors is not uniform, and the images in different exploration areas may present completely different amplitude distribution, it is common to see the weak and strong events in different positions of the same seismic image. If the amplitudes of events in training datasets are similar, the trained SIST which contains lots of W-MSA and SW-MSA blocks will calculated the SA among samples using the similar weights within a window, and then the amplitudes of weak events will be ignored or are anomaly raised to the same amplitude value as strong events. Therefore, for improving the reconstruction of weak events and preservation of amplitudes, we make half of the 2D slices introduced above as mixture of strong and weak events. We multiply the amplitudes of events in a 2D slice with a random coefficient which is selected between 0.3~0.8, and the multiplication ranges are in depth (or time) direction with a random width which is selected between 20%~50% continues samples of all the samples.

**Model training and evaluating**

*Training process*

For demonstrating the feasibility of the proposed SIST, we use the classic U-Net (Ronneberger et al., 2015) for comparison which is widely used for many geophysical tasks. The Adam algorithm is used as optimizer. Table 1 lists the specific settings of the hyper-parameters. The learning rate decays at the $230^{th}$, $270^{th}$ and $300^{th}$ epoch. The predictions of networks can be expresses as

$$\hat{y} = Net(y;\omega,b), \qquad (9)$$

where y and $\hat{y}$ represent the input data and prediction, respectively, while $\omega$ and $b$ denote the weights and biases of networks which are updated during training. The solution of $L_1$ loss function is with better sparsity, which is suitable for the division of stratigraphic interfaces in seismic images and improve the anti-noise capability of SR methods as well. In addition, the structural similarity (SSIM) measures the similarity between two images in structures, especially the multiscale structural similarity (MS-SSIM) (Wang et al., 2003) by down-sampling the original data several times to obtain multi-scale images to pay more attention to the local structure perturbations which is beneficial in SR tasks to improve the resolution of local events. Thus, we use a mixed loss function to measure the distances between input data and prediction which is proved a better performance in image restoration (Zhao et al., 2016)

$$L(\omega, b) = \varphi L_1 + (1-\varphi) L_{MS-SSIM}, \tag{10}$$

where $L$ represents the loss function used for SIST, and $L_{MS\text{-}SSIM}$ represents the loss function of MS-SSIM. $\varphi$ represents the tradeoff between $L_1$ and $L_{MS\text{-}SSIM}$. After many tests, it is concluded that the SR effect is the best when $\varphi$ equals 0.6. The $L_1$ loss function can be expressed as

$$L_1(\hat{y}, g) = \frac{1}{N} \sum_{n=1}^{N} |\hat{y}_n - g_n|, \tag{11}$$

where $g$ and $N$ represent the ground truth and the total number of samples, respectively. The loss function of MS-SSIM can be expressed as

$$L_{MS-SSIM}(\hat{y}, g) = \left[l_M(\hat{y}, g)\right]^{\alpha_M} \prod_{m=1}^{M} \left[c_m(\hat{y}, g)\right]^{\beta_m} \left[s_m(\hat{y}, g)\right]^{\gamma_m}, \tag{12}$$

where

$$l(\hat{y}, g) = \frac{2\mu_{\hat{y}} \mu_g + c_1}{\mu_{\hat{y}}^2 + \mu_g^2 + c_1},$$

$$c(\hat{y}, g) = \frac{2\sigma_{\hat{y}}\sigma_g + c_2}{\sigma_{\hat{y}}^2 + \sigma_g^2 + c_2}, \tag{13}$$

$$s(\hat{y}, g) = \frac{Cov_{\hat{y}g} + c_3}{\sigma_{\hat{y}}\sigma_g + c_3},$$

where $M$ is the total number of scales, and we set $M$ equals 5 in this paper. $l$, $c$ and $s$ represents the measurement of luminance, contrast, and structure, respectively. $α$, $β$ and $γ$ represent the weights of $l$, $c$ and $s$, respectively, and the value of $α$, $β$ and $γ$ should be positive. $μ$ and $σ$ represents the mean and standard deviation of $g$ and $\hat{y}$. $Cov$ denotes the covariance between $g$ and $\hat{y}$. $c_1$, $c_2$ and $c_3$ are constants to keep the denominator from being 0. Especially, when $α$, $β$ and $γ$ are all equal to 1, and $c_3$ equals to half of $c_2$, equation (12) becomes the commonly used SSIM (Wang et al., 2004).

Table1. Hyper-parameters of networks

| Hyper-parameters | SIST | U-Net |
| --- | --- | --- |
| Network depth | 156 | 65 |
| Size of input data for training/ testing | 128×128/ not limited | 128×128, not limited |
| Size of output data for training /testing | 256×256/ double size of the input data | 256×256/ double size of the input data |
| Batch size | 8 | 8 |
| Learning rate | [$2^{-4}$, $1^{-4}$, $5^{-5}$] | [$2^{-4}$, $1^{-4}$, $5^{-5}$] |
| Number of epochs | 350 | 350 |
| Size/ stride of windows | 8×8/ 4 | / |
| Size/ stride of filters | 1×1, 3×3/ 1 | 3×3/ 1 |
| Pairs of data for training/ validation/ testing | 2400/300/300 | 2400/300/300 |

Training is conducted on a computer workstation with the following configurations: 12 vCPU Intel Xeon Platinum 8352V CPU @ 2.10GHz, NVIDIA RTX 4090 GPU with 24 G RAM, and the used deep learning framework is Pytorch V2.0.0 based on Python V3.8. In addition, each network is trained 10 times, and evaluate the accuracy of predictions to select the optimal models.

*Evaluating methods*

For quantitatively evaluating the trained models to select the optimal models and compare the behaviors of different methods, we use root mean square error (RMSE) to evaluate the amplitude preservation, peak signal-to-noise ratio (PSNR) to evaluate the denoising capability, and SSIM to evaluate the reconstruction of geological structures. RMSE can be calculate by

$$\text{RMSE}(\hat{y}, g) = \sqrt{\frac{\sum_{n=1}^{N}(g-\hat{y})^2}{N}}, \tag{14}$$

and PSNR can be calculated by

$$\text{PSNR}(\hat{y}, g) = 10\log_{10}\frac{\text{MAX}^2(g)}{\text{RMSE}^2(\hat{y}, g)}, \tag{15}$$

where MAX(·) represents take the maximal absolute value of the data in bracket. In addition, the maps of local SNR and local SSIM are drawn to show the performance of different methods further. However, filed data which lacks the ground truth is hard to be evaluated quantitatively using the index introduced above. Thus, we use the comparisons of 2D predicted images, 1-dimensional (1D) traces and 1D frequency spectra to evaluate the predictions of field data.

## NUMERICAL TESTS

In this section, we first demonstrate the superior performance of SIST on synthetic data which contains the testing datasets and the image obtained from Marmousi model. Then we compare the behaviors of SIST and U-Net on field data collected from Netherlands F3 block and New Zealand Parihaka 3D survey. The field data with high and low SNR, and the inline and crossline profiles in 3D volume with different sizes are all tested in this section.

**Tests on synthetic data**

*Test on testing data*

10% of the generated synthetic data which are not participate in training are used for testing. Figure 2 shows one of the tests. Figure 2a and 2b show a pair of synthetic LR and HR seismic images which are the input data and ground truth (GT) of the networks, respectively. Figure 2c shows the prediction of the trained U-Net, which is generally similar to the GT. However, there are many details deviate from the GT, such as the artifacts are introduced between layers and the reconstructed amplitudes are not accurate enough. The prediction of SIST is shown in Figure 2d, which is similar to the GT both in general and details. For detailed comparison, the red arrows in Figure 2 indicates the large fault in the prediction of U-Net is blurred, but it is clear in the prediction of SIST. There are obvious artifacts exist in the prediction of U-Net shown in the yellow box where is the area of folds with large dips, however, SIST behaves better and both the shapes and amplitudes are correct. Especially, the reconstruction of tiny fault (pointed by yellow arrows) is accurate by SIST which is totally disappeared in the prediction of U-Net. Table 2 shows the quantitative comparison of predictions obtained by different networks. We can see that the RMSE of the prediction obtained by SIST is close to 0 and it is much smaller than that of the U-Net which demonstrates the amplitude preservation of SIST. The PSNR of the prediction obtained by SIST is large and it is almost double of the prediction obtained by U-net which indicates SIST is with a superior capability of denoising. Besides, the SSIM of the prediction obtained by SIST is close to 1 and it is much larger than that of the prediction obtained by U-Net proves that SIST with global receptive field can better reconstruct the shapes of geological structures in space.

**Table 2. Quantitative comparison of predictions obtained by different networks**

|  | SIST | U-Net |
|---|---|---|
| RMSE | 0.0564 | 0.2892 |
| PSNR | 28.4962 | 14.2979 |
| SSIM | 0.9576 | 0.3752 |

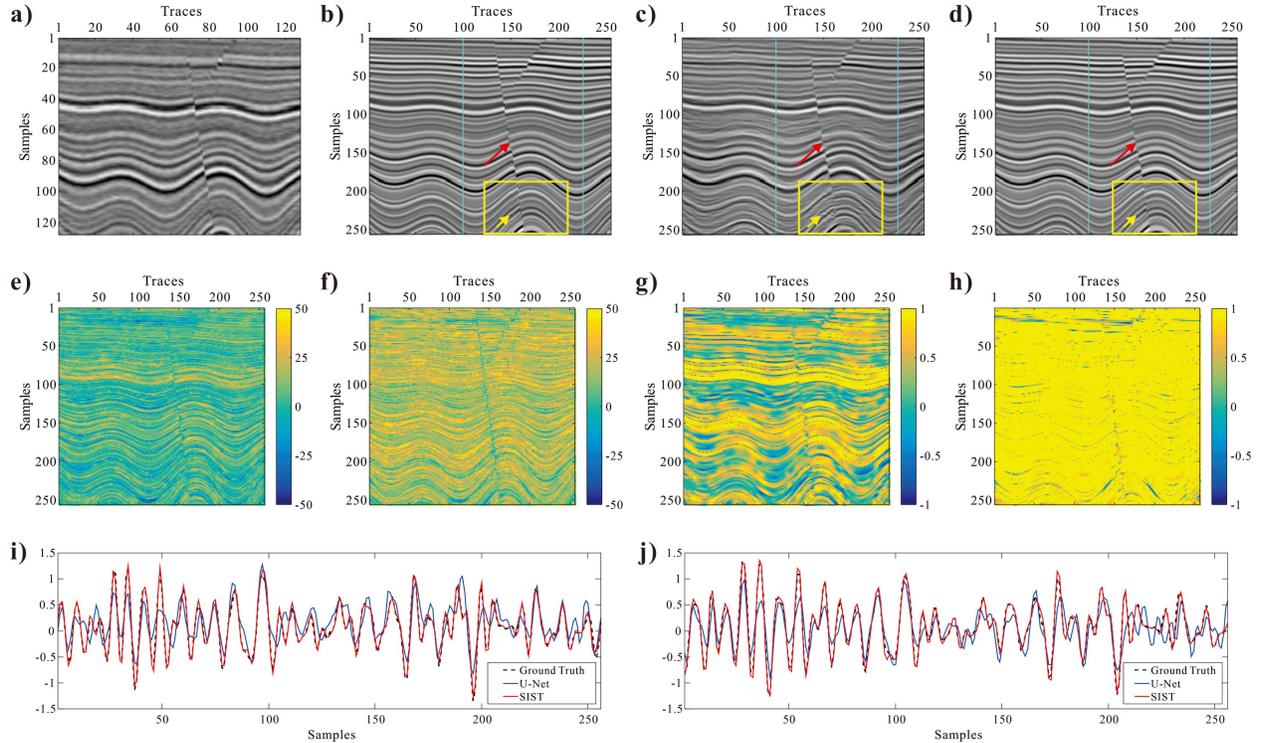

**Figure 2.** Comparisons of SR results by different methods on testing data: (a) the LR input data, (b) the ground truth, (c) the prediction of U-Net, (d) the prediction of SIST, (e) and (f) are the local SNR of the predictions obtained by U-Net and SIST, respectively, (g) and (f) are the local SSIM of the predictions obtained by U-Net and SIST, respectively, (i) and (j) are the comparisons among 100$^{th}$ and 225$^{th}$ trace of the data shown in (b), (c) and (d), respectively.

For comparing the performance of SIST and U-Net further, we plot local SNR (Figure 2e and 2f), local SSIM (Figure 2g and 2h) and single traces (Figure 2i and 2j), respectively. From Figure 2e and 2f we can see that the local SNR in most of the areas of the prediction obtained by U-Net is low and some of the values are negative, but the local SNR of the prediction obtained by SIST is high and only little values are close to 0. From Figure 2g we can see that although the local

SSIM in some areas of the prediction obtained by U-Net is close to1, the values in other areas are low and even negative. However, the local SSIM of the prediction obtained by SIST show most of the values are 1 (Figure 2h). Then we extract the 100$^{th}$ and 225$^{th}$ trace from the GT, prediction by U-Net and SIST (indicated by blue lines), respectively. Both the comparisons among single traces shown in Figure 2i and 2j indicate the amplitude of the prediction obtained by SIST fits that of the GT well, but the differences between the amplitudes of prediction obtained by U-Net and GT are large, and pseudo events are introduced. Therefore, the test on testing data demonstrates the feasibility of SIST preliminarily.

*Test on the seismic image obtained from Marmousi model*

Marmousi model is a classically geological model which is widely used to test methods on modeling of wave equations, inversion and migration. The seismic image obtained from Marmousi model is more similar to the features of field stratigraphic structures than the seismic images which contains only simple geological structures and texture features in the training and testing datasets. We use Marmousi velocity model and a constant density model with the same size of Marmousi model to generate an reflection coefficient model, and then convolve the Ricker wavelet with the reflection coefficient model to obtain the corresponding seismic images. Similar to the generation process of training datasets, we use the Ricker wavelet with a peak frequency of 15 Hz and 50 Hz to generate the LR and HR images, respectively. Besides, we add colored noises to the LR image.

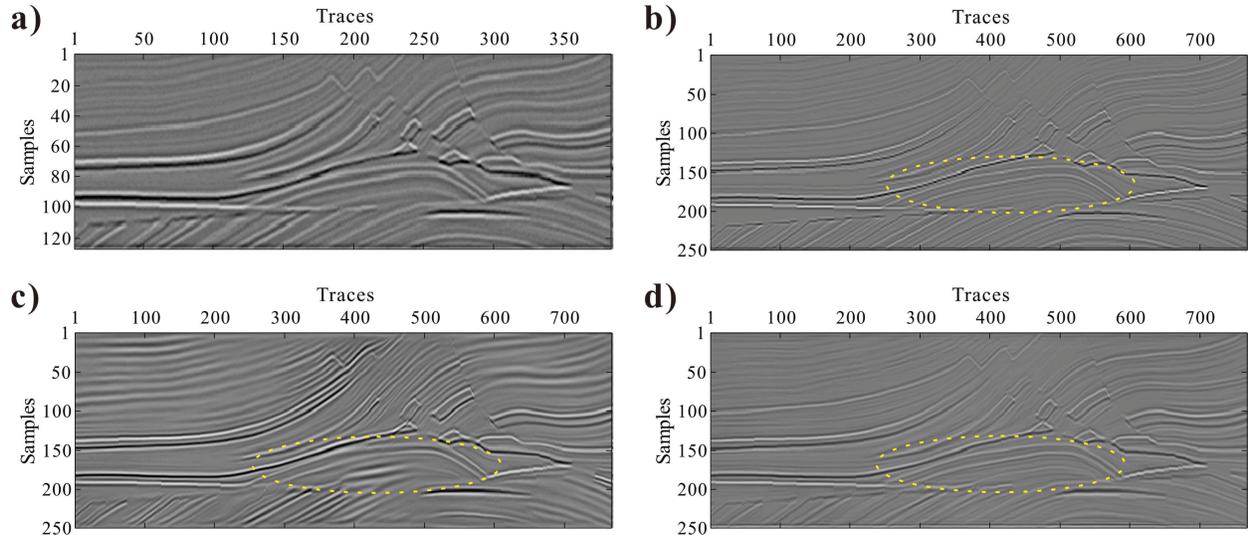

**Figure 3.** Comparisons of SR results by different methods on seismic image obtained from Marmousi model: (a) the LR input data, (b) the ground truth, (c) the prediction of U-Net, (d) the prediction of SIST.

Figure 3a and 3b show the LR and HR images, respectively. The events in the LR image are coarser and many details are covered up such as thin layers and weak reflectors, and the geological structures are blurred compared to the HR image. Figure 3c shows the prediction of U-Net, which shows an enhance of resolution in some extent. Nonetheless, there are incorrect amplitudes and insufficient peak frequency extrapolation exist in the image. The faults are not clear and artifacts appear in many areas. In addition, the anticline (indicated by yellow circles) which is an important target is blurred and almost nothing can be distinguished. However, SIST behaves much better than U-Net, and the image predicted by SIST (Figure 3d) is very close to the GT. No matter the shapes of geological structures or the amplitudes of events especially weak events and thin layers, the prediction of SIST fits the GT well, and there is no artifacts introduced. Thus, this test demonstrates the potential of SIST on applying to the SR of complex seismic images.

**Testing on field data**

*SR for the seismic images from Netherlands F3 block*

The Netherlands F3 block collected 3D marine seismic data, and we select two patches of seismic images for testing.

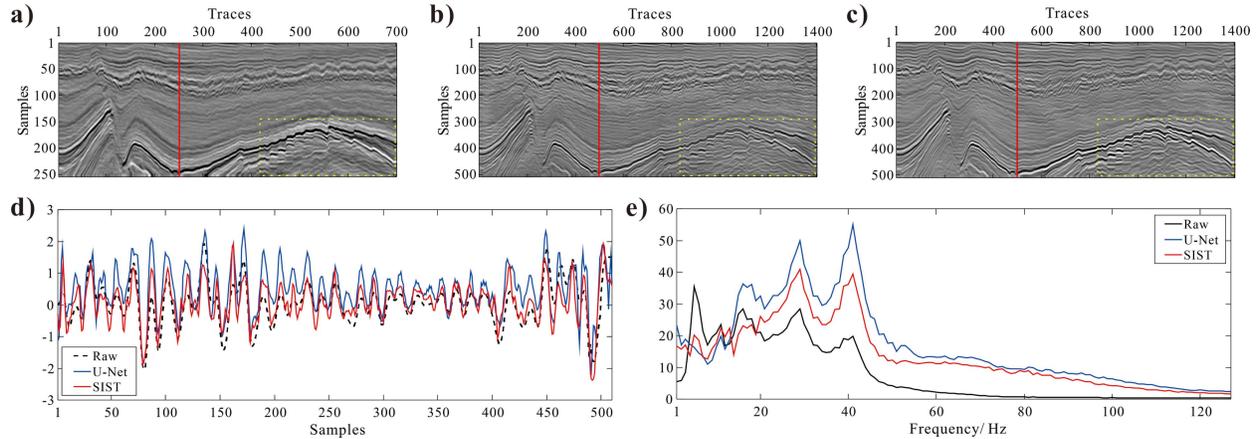

**Figure 4. Comparisons of SR results by different methods on the first patch of data from F3 block: (a) the raw data, (b) the prediction of U-Net, (c) the prediction of SIST, (d) comparison among single traces with 250$^{th}$ trace from (a) and 500$^{th}$ trace from (b) and (c), (e) comparison among average amplitude spectra of the data shown in (a), (b) and (c).**

The first patch is shown in Figure 4a which is an image with high SNR. The prediction of U-Net and SIST are shown in Figure 4b and 4c, respectively. Both the predictions improve the resolution of the horizons, folds and large faults in the raw data, and there is almost no artifact introduced. For more detailed comparison, the images in yellow boxes indicate that the numerous small faults predicted by SIST are clearer with a higher resolution, and the recovered weak events below the strong reflectors is clearer and more continuous. Then we compare single traces with the 250$^{th}$ trace from the raw data and 500$^{th}$ trace from the predictions (Figure 4d), and the selected traces are indicated by red lines. We perform linear interpolation on the raw data to make its number of samples consistent with the predictions. The overall amplitude variation trends of the predictions by U-Net and SIST are similar to that of the raw data, and the amplitudes are enhanced to improve resolution where the raw data is less volatile. However, the absolute values of the amplitudes predicted by U-Net are too large which destroys the true amplitude of events, but the amplitude preservation of SIST is good. Then we compare the average amplitude spectra of the data shown in Figure 4a, 4b and 4c. Both the spectra of the prediction by U-Net and SIST enhance the peak frequency a lot, while maintaining the low frequencies of the raw data is also well. Low frequency consistency ensures that the overall variation trends of formations are consistent with that of the raw data, and richer high frequencies provide more details for improving resolution.

The second patch is shown in Figure 5a, and the SNR of this patch is lower than that of the first patch. The prediction of U-Net and SIST are shown in Figure 5b and 5c, respectively. Similar to the first patch, both the predictions improve the resolution of the raw data. However, due to the noises contaminate the raw data in the range of 300~400 samples where are mostly weak events, there are a lot of fake events appear in the prediction of U-Net, and as a contrast, SIST show strong capability of denoising which recovers the weak events from strong noises and does not introduce any artifacts (indicated by yellow circles). As indicated by yellow boxes, the several small faults predicted by U-Net are blurred, and the stratigraphic continuity on both sides of the faults is poor, however the prediction of SIST show clear faults and the stratigraphic interfaces on both sides of the faults are clearer as well. In addition, two tiny conical faults reconstructed by U-Net is not clear, which displays the same strata in the hanging wall and foot wall of the left fault are connected together, however the prediction of SIST show clear interfaces on both sides of the conical faults (indicated by red arrows). Figure 5d shows the amplitude preservation of SIST is better than that of U-Net by comparison among single traces which are selected the $150^{th}$ trace from raw data and $300^{th}$ trace from predictions (indicated by red lines). Figure 5e shows both U-Net and SIST enhance the peak frequency of raw data a lot and keep consistent with the low frequencies of the raw data.

The SR tests on seismic images from Netherlands F3 block demonstrate the generalization and anti-noise capability of SIST.

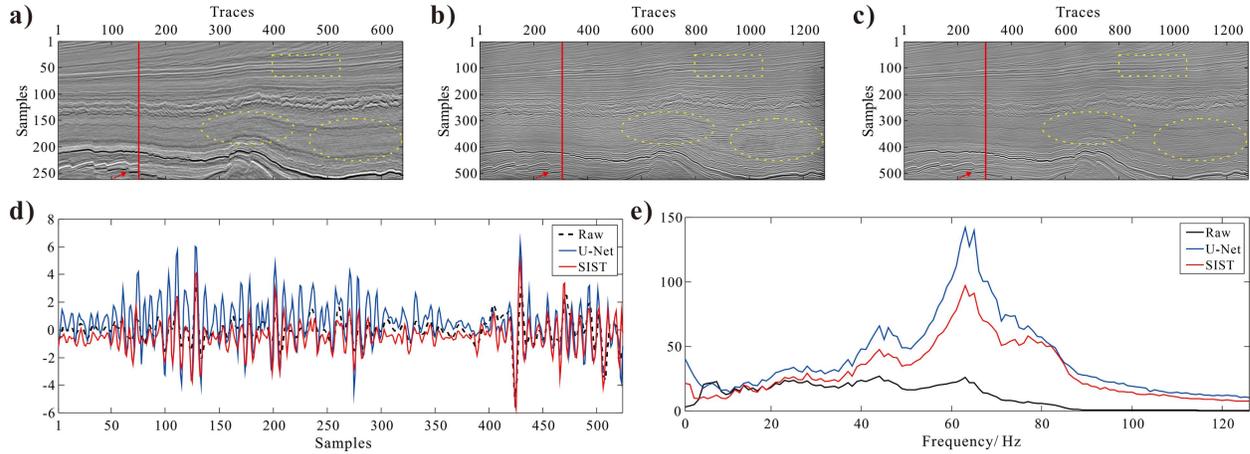

**Figure 5. Comparisons of SR results by different methods on the second patch of data from F3 block: (a) the raw data, (b) the prediction of U-Net, (c) the prediction of SIST, (d) comparison among single traces with 150th trace from (a) and 300th trace from (b) and (c), (e) comparison among average amplitude spectra of the data shown in (a), (b) and (c).**

*SR for the seismic images from New Zealand Parihaka 3D survey*

For demonstrating the generalization and SR capability of the proposed method further, we apply the trained model on field data from New Zealand Parihaka 3D survey. We select the 1376th inline and 4025th crossline profiles for testing.

Figure 6a shows the raw 1376th inline profile. The shallow parts of the image is dominated by the horizons with little dips and high SNR which makes both U-Net and SIST can obtain good SR results for these parts of the image. Therefore, the image in the yellow box shown in Figure 6a (enlarged in Figure 6b) with complex features is selected for comparison. There are folds, little faults and horizons with large dips in Figure 6b, and the SNR is low in the lower part of this image. Figure 6c and 6d show the predictions of U-Net and SIST, respectively. It can be seen from Figure 6c, the contrast of the whole image is abnormally uneven, which visually shows that some parts of the image are bright and some parts are dark. This phenomenon indicates that the relative amplitude preservation of U-Net becomes poorer while predicting more complex images. However, the contrast of the prediction by SIST is even, which maintains the normalized energy distribution of the raw data. Influenced by strong noises, the predicted events by U-Net from low SNR raw data are still blurred, and there is almost no improvement of resolution, however, SIST recover the clear and continuous events from noises especially the weak events (indicated by yellow circles). The prediction accuracy of SIST for tiny scatterers and the events around them is also higher than that of U-Net (indicated by red circles). Besides, for the small faults and horizons with weak amplitudes indicated by red arrows, the prediction of SIST are much better than those of U-Net in terms of the clarity of fault interfaces and the continuity of horizons. In addition to the marked parts of the images we introduced above, there are many areas between the predictions where SIST show similarly superior performance as well.

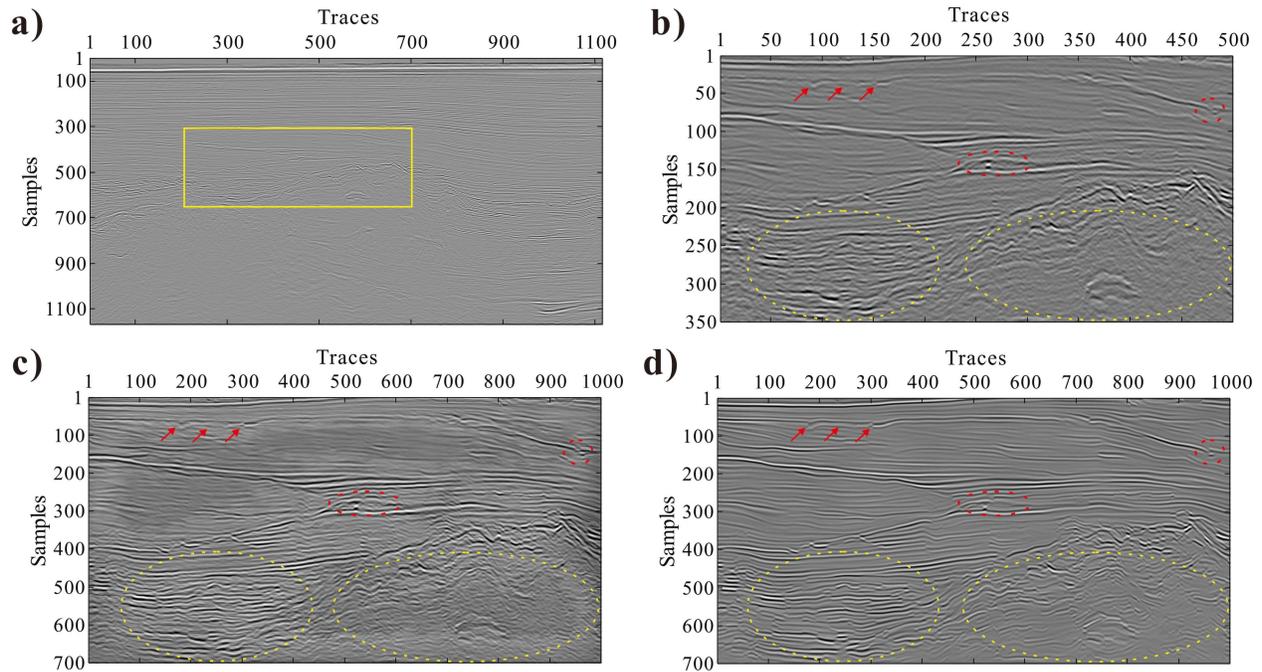

**Figure 6. Comparisons of SR results by different methods on the inline profile from Parihaka 3D survey: (a) the raw 1376$^{th}$ inline profile, (b) the enlarged image in the yellow box shown in (a), (c) the prediction of U-Net, (d) the prediction of SIST.**

Then we test a crossline profile which are different from the inline profile in characteristics, and we select the 4025$^{th}$ crossline profile for testing (Figure 7a). There are more faults including little and large faults exist in this profile. The shallow parts of the image is also dominated by the horizons with little dips and high SNR. Therefore, the image in the yellow box shown in Figure 7a (enlarged in Figure 7b) with complex features is selected for comparison. Figure 7c and 7d show the predictions of U-Net and SIST, respectively. Similar to Figure 6c, the contrast of the whole image shown in Figure 7c is also uneven, which indicates that the amplitude preservation of U-Net is poor. However, the performance of SIST remains stable, and the prediction consistent with the contrast of the raw data can still be obtained by SIST while processing data with different characteristics. The area indicated by yellow triangles is the stratum sandwiched between two large faults, and this part of the image is with low SNR and the amplitudes of events are weak. The prediction by U-Net in this area is still blurred, and the amplitudes of weak events are even weaker than that of the raw data. However, SIST obtains a better predicted results in this area which shows clearer shapes of stratum and obvious boundaries with the faults on both sides. The yellow circles show areas with multiple faults. Although the prediction by U-Net appears clearer faults in this area, the events are thicker which means the peak frequencies are lower, and the resolution enhancement is limited. However, SIST predicts thinner events in this area and the faults are clearer than that of the prediction by U-Net. In addition, there are multiple small faults positioned longitudinally which are indicated by red arrows, and the SNR in this area is low. Affected by noises, the faults in the prediction by U-Net are still unclear, while SIST recovers multiple small faults with high resolution from strong noises. Similarly, in addition to the marked parts of the images we introduced above, there are many areas between the predictions where SIST show similarly superior performance as well.

The SR tests on inline and corssline profiles from 3D survey show the potential of SIST on processing complex 3D field data. In addition, SIST show strong generalization and superior performance on SR tasks.

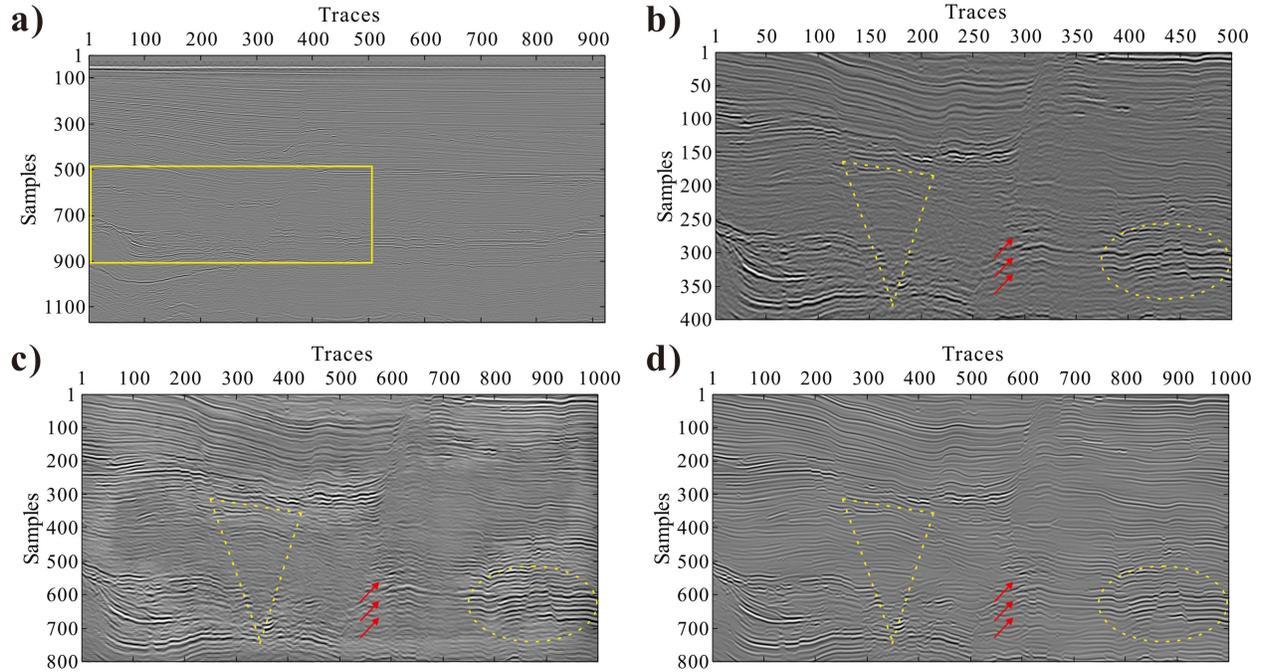

Figure 7. Comparisons of SR results by different methods on the crossline profile from Parihaka 3D survey: (a) the raw 4025[th] crossline profile, (b) the enlarged image in the yellow box shown in (a), (c) the prediction of U-Net, (d) the prediction of SIST.

## DISCUSSIONS

In this section, we discuss the relationship between amplitude preservation as well as recovery of weak events and the architecture of SIST. In the section of methods, we introduced that the additional edge images and the directly up-sampling to the raw input data aims at providing more information of true amplitudes and increasing the focus on weak events. Thus, we perform a ablation test to demonstrate the the validity of these two operations. The post-stack profiles of North Sea Volve field are suitable for this test which is shown in Figure 8a. The SNR of the entire image is low, and the amplitudes of events in upper and lower parts of the image are strong, while that of the middle part are weak but contains many geological structures. The peak frequencies of the events in upper and lower parts of the image are high and low, respectively. In order to show the SR results more clearly, we select the image in the yellow box shown in Figure 8a (enlarged in Figure 8b) for comparison. The characteristics of the image shown in Figure 8b is a microcosm of the entire profile shown in Figure 8a, which contains events with different amplitudes, peak frequencies and the SNR is low as well.

Figure 8c and 8d show the predictions of ablative SIST and SIST, respectively. The ablative SIST means the edge input and up-sample to the raw data are removed from SIST. It is obvious in Figure 8c that the overall amplitudes are weak, and the amplitudes of some weak events are even weaker than that of the raw data shown in Figure 8b. However, the overall contrast of the image shown in Figure 8d is even, and the events are clear and continuous. The weak events with low frequencies predicted by ablative SIST are even more blurred than that of the raw data, but SIST achieves recovering the weak events clearer and more continuous (indicated by yellow circles). Especially, the interfaces of small faults are clearer under the SR by SIST (indicated by yellow arrows). The images in the red boxes are weak events including horizons, folds and faults with decreasing peak frequency from shallow to deep, and the contamination of noises is severe. The prediction by ablative SIST of the image in the red box is worse than that of the raw data, and the geological structures are hardly to be distinguished. As a contrast, the recovered events by SIST are clear for each kind of geological structure, and the noises are suppressed well. Besides the comparisons on weak events, we compare the SR results on an area with relatively strong events and contains multiple faults (indicated by yellow boxes shown in Figure 8b, 8c and 8d). Although the prediction by ablative SIST show enhancement of resolution in some extent, the resolution is still lower than that of the prediction by SIST, which indicates the edge input and up-sample to the raw data also work well on improving the resolution of strong events.

Then we compare single traces with the 50th trace from the raw data and 100th trace from the predictions (Figure 8e), and the selected traces are indicated by blue lines. The overall amplitude variation trends of the prediction by SIST are similar to that of the raw data, but the amplitudes of the prediction by ablative SIST are obviously small in general, especially for the weak events between 250~400 samples. The average amplitude spectra of the data are shown in Figure 8f. Both the spectra of the predictions by ablative SIST and SIST enhance the peak frequency, but the enhancement of ablative SIST is limited. In addition, the maintenance of low frequencies to the raw data by ablative SIST is worse than that of SIST, which destroys the amplitude variation trends of the raw data.

By this ablation test, we demonstrate the effectiveness of the edge input and up-sample to the raw data in amplitude preservation and recovery of weak events.

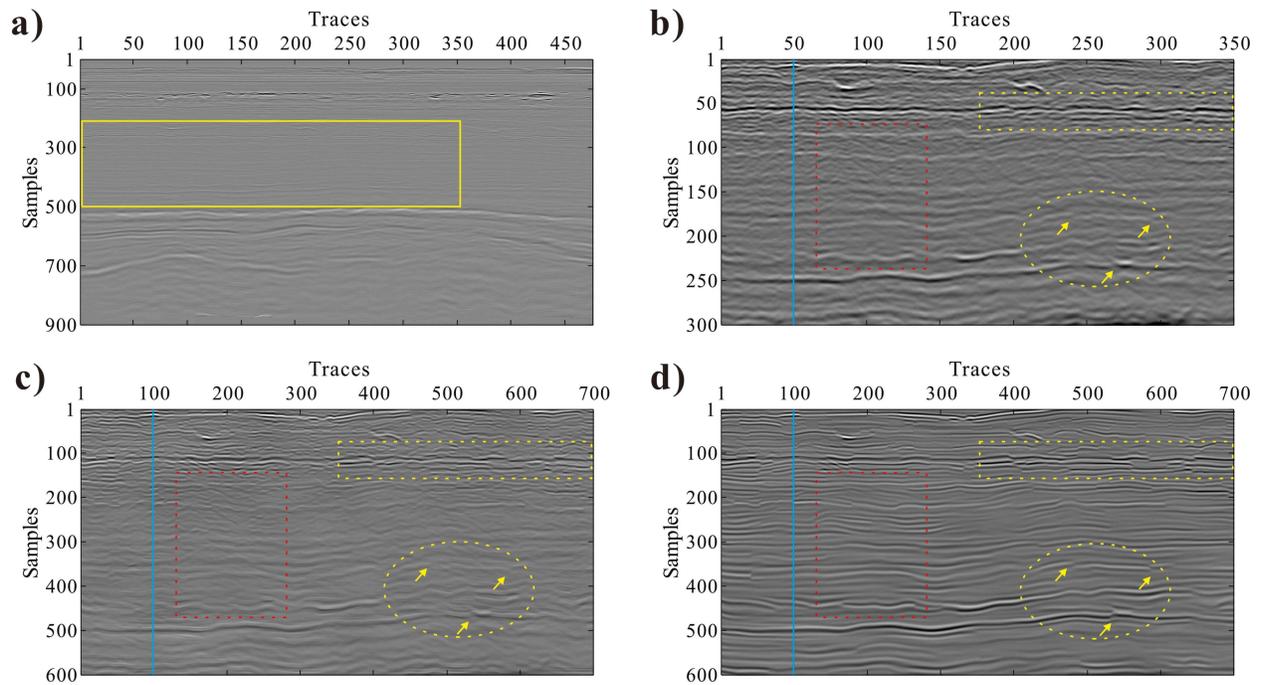

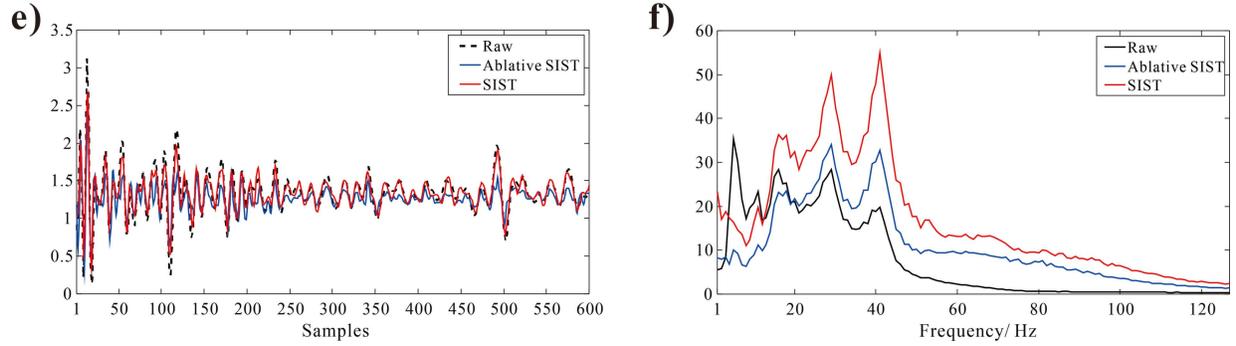

**Figure 8.** Ablation test on field data from North Sea Volve field: (a) the entire raw data, (b) the enlarged image in the yellow box shown in (a), (c) the prediction of U-Net, (d) the prediction of SIST, (e) comparison among single traces with 50$^{th}$ trace from (b) and 100$^{th}$ trace from (c) and (d), (f) comparison among average amplitude spectra of the data shown in (b), (c) and (d).

## CONCLUSIONS

In this study, we propose a novel deep learning based model named SIST to achieve state of the art SR results for seismic images. SIST uses Swin-Transformer as the backbone to extract and fuse the local and global features of images, which not only pays attention to the amplitudes and SNR of local events, but also ensures the correctness of the shapes in space of geological structures. We extract the edge information with true amplitude as an additional input channel, and up-sample the raw data directly to constrain the output of SIST to improve the amplitude preservation and recovery of weak events. The synthetic training datasets with similar features to field seismic images is generated for supervised learning, which ensures the prediction accuracy and generalization of SIST. Numerical tests on the testing datasets, synthetic image from Marmousi model, field data obtained from different exploration regions with complex features including low SNR, mixture of weak and strong events, mixture of events with different peak frequencies, inline and crossline profiles from 3D survey, and multiple geological structures, demonstrate the strong generalization, amplitude preservation, recovery of weak events and denoising capability of the proposed method. The proposed method will help to improve the accuracy of geological interpretation, and can be applied to the data of old exploration areas to discover new geological targets. In the future, we will keep researching on this method to make it directly perform SR tasks on 3D volumes to improve the continuity of prediction results between different slices.